\documentclass[10pt,twoside]{article}

\usepackage{8OSME-style}

\title{An Efficient Enumeration of Flat-Foldings : Study on Random Single Vertex Origami 
}
\authorlist[Nakajima]{Chihiro Nakajima} 
\affiliations{Chihiro Nakajima \at Faculty of engineering, Tohoku Bunka Gakuen University, 6-45-1 Sendai Miyagi, Japan, \email{chihiro.nakajima@ait.tbgu.ac.jp}}

\makeatletter
  \let\runtitle\@title
  \let\runauthor\shortauthor
\makeatother

\begin{document}

\maketitle

\begin{abstract}
This paper deals with themes such as approximate counting/evaluation of the total number of flat-foldings for random origami diagrams, evaluation of the values averaged over various instances, obtaining forcing sets for general origami diagrams, and evaluation of average computational complexity. An approach to the above problems using a physical model and an efficient size reduction method for them is proposed.
Using a statistical mechanics model and a numerical method of approximate enumeration based on it, we give the result of approximate enumeration of the total number of flat-foldings of single-vertex origami diagram with random width of angles gathering anound the central vertex, and obtain its size dependence for an asymptotic prediction towards the limit of infinite size.\\
In addition, an outlook with respect to the chained determination of local stacking orders of facets caused by the constraint that prohibits the penetration of them is also provided from the viewpoint of organizing the terms included in the physical model.
A method to efficiently solve the problem of the determination or enumeration of flat-foldings is discussed based on the above perspectives. 
This is thought to be closely related to forcing sets.
\end{abstract}

\section{Introduction}
\label{sec:introduction}
%

Statistical mechanics deals with configurations given by combinations of variables that take on a small number of states (for example, binary variables)\cite{Ising}.
It introduces realization probabilities to configurations and discusses the properties of moments and large deviation functions, and is closely related to mathematics through probability theory and combinatorics.
In mathematical and computer science research on origami, a statistical mechanics perspective has been introduced to some problems (though it is sporadically, not based on a unified perspective)\cite{Hull_t,Assis}.

\section{Physical model for folding of origami diagram}
\label{sec:physmodel}
%
In this study, the folding diagrams are generated according to probabilistically generating algorithm explained below.
With the total number of facets which is descried as $n$ is fixed, large numbers of instances are generated by randomly giving widths of angles around the center, and discuss the statistical properties of the resulting set of instances.
We try to obtain asymptotic predictions on the behavior in the limit of an infinite number of facets $n \to \infty$ from the sequences of those with finite number of facets. 

\begin{SCfigure}[][ht]
  \centering
  \includegraphics[width=50mm]{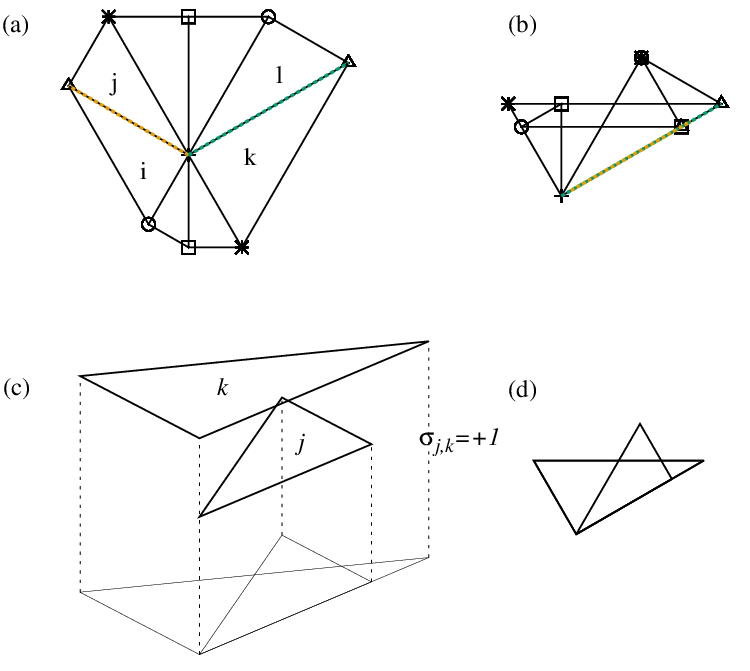}
  \caption{(a)Example of origami diagram. Each edge in the figure represents a crease. In this figure there are no overlaps of facets. (b)Corresponding pre-folded diagram, which describes the overlaps of facets when the figure (a) is folded along the creases. Each vertiex indicated by the same mark is the same as that in Fig. (a). (c)Schematic picture of introduction of the Ising variable to a local layer-ordering.\label{fig:fig1}}
  \label{fig:fig1}
\end{SCfigure}

Attempts to introduce physical models into origami have been made in the past.
It includes one that deals with geometric constraints on the bending of flat structures in three-dimensional space\cite{FGM},
 one aimed at the physical properties of polymers and membranes\cite{NPW},
 one that extracts the phase transition phenomenon seen in self-folding origami\cite{Assis},
 and one that uses it in the context of mathematical research\cite{Hull_t}.

In this paper we consider the model consisting of binary variables $s_{i,j}\in \{-1,+1\}$ with the condition subjected with the form of products of two or four variables\cite{CN1} to traet the stacking of facets and determinate flat foldability accurately.
The variables $s_{i,j}$ represent the vertical relationship in stacking of two facets $i$ and $j$ in the origami diagram.
Each realization of $s_{i,j}\in \{-1,+1\}$ represent the global stacking order of $n$ pieces of facets.\\
The conditions to prohibit the situation that a crease is penetrated or intruded by other facets are imposed with the following terms in the energy function,
\begin{align}
  E^{(i)}_{i,j;k}
  &=\frac{1}{2}\big(1-J_{(ik)(kj)}s_{i,k}s_{k,j}\big)\qquad\qquad\qquad\qquad \label{eq:term_J}\\
  E^{(c)}_{i,j,k}
  &=\frac{1}{4}\big(1-L_{(ij)(jk)}s_{i,j}s_{j,k}-L_{(jk)(ki)}s_{j,k}s_{k,i}-L_{(ki)(ij)}s_{k,i}s_{i,j}\big), \label{eq:terms_L}\\
  E^{(q)}_{i,j ; k,l}
  &=\frac{1}{2}\Big(1-K_{ijkl}s_{i,k}s_{i,l}s_{j,k}s_{j,l}\Big),\label{eq:terms_K}
\end{align}
where $J_{(ik)(kj)}=-\tau_{ik}\tau_{kj}$,
$K_{ijkl}=\tau_{ik}\tau_{il}\tau_{jk}\tau_{jl}$,
$L_{(ij)(jk)}=-\tau_{ij}\tau_{jk}$,
and $\tau_{ij}$ is the sign of the difference between two indices of facets, namely $\tau_{ij}=\textit{sign}(j-i)$.
Note that for the variable $s_{i,j}$, pay attention to the order in which subscripts are written, is assigned only when $i<j$ and vice versa.
The term (\ref{eq:term_J}), which prohibits a facet $k$ from intruding between two facets $i$ and $j$ which are connected by a crease, is assigned for a geometry where a crease with $(i,j)$ has an overlap with a facet $k$ in the pre-folded diagram.
The term (\ref{eq:terms_L}), which prohibits cyclic stacking among the three facets $i$, $j$, $k$, is assigned for a geometry where three facets $i,j$ and $k$ simultaenously share an area.
The term (\ref{eq:terms_K}), which prohibits the unrealizable overlap between the four facets $i$, $j$, $k$, $l$ that form two creases in (sometimes partially) coincident position, is assigned for a geometry where the two creases each consisting two facets $i$ and $j$, $k$ and $l$ with respect are in coincident position.
The terms with the form (\ref{eq:term_J})-(\ref{eq:terms_K}) are respectively assigned to each geometries of overlap of a crease and a facet, simultaneously shared area, and coincident creases in the pre-folded diagram.
Hence, the total energy function is described as followings,
\begin{equation}
H(\{s\})=\sum_{(i,j ; k)}E^{(i)}_{i,j;k} + \sum_{(i,j,k)}E^{(c)}_{i,j,k} + \sum_{(i,j;k,l)}E^{(q)}_{i,j;k,l}, \label{eq:tot_en_func}
\end{equation}
where the each summation is taken over all corresponding geometries in the pre-folded diagram.
Thus the problem of finding the flat folding is casted as an optimization problem,
that of finding the conbination of $\{s_{i,j}\}$ with $H=0$.

\section{Single-vertex origami diagram}
\label{sec:sec3}
Preparing a flat-foldable diagram in general is considered to be equivalent to a combinatorial optimization problem itself.
In fact, when an origami diagrams are generated at random angles, the frequency of the flat-foldable diagrams is extremely small, so that it is almost unrealistic to discuss the foldability problem.

Instead of the generation of general diagrams, generating ones of single-vertex structure such as exhibited in Fig.\ref{fig:fig1}(a) is performed in this research.
The properties of single-vertex diagrams have been actively researched in the context of determining foldability under a fixed crease pattern (mauntain-valley assignment), enumerating foldable crease patterns, and enumerating flat-foldings (global stacking-order of facets)\cite{Hull_t2}.
It is known that the foldability of a single vertex diagram can be determined using the Kawasaki's theorem\cite{Kaw_s}.
This themrem makes to generate foldable diagrams with random angles around the center possible by maintaining the condition that the alternating sum of those is $0$.
Regarding the enumeration of the total number of flat foldings at least for the case of random angles, even in single vertex diagrams there is still no known conclusion, as far as the narrow knowledge of the auther.

\subsection{Numerical procedure for generating each instance}
\label{subsec:ForInstances}
%
For an origami diagram whose total number of facets is $n$, the total number of pairs of facets is $n(n-1)/2$, so the upper limit of $N$, the number of pairs which the local layering order should be considered, is also this value. 
However, depending on the instance, there are facets that do not have an overlapping area in the folded diagram. 
No variables are assigned to these pairs, as there is no need to consider direct hierarchical relationships. 
As a result, the value of $N$ for the origami diagram of $n$ facets roughly takes a value close to the upper limit, but takes various values depending on the details of the overlap.
Thus the number of variables $s_{i,j}$ for the corresponding optimization problem is given by $N$.

\begin{SCfigure}[][ht]
\includegraphics[width=50mm]{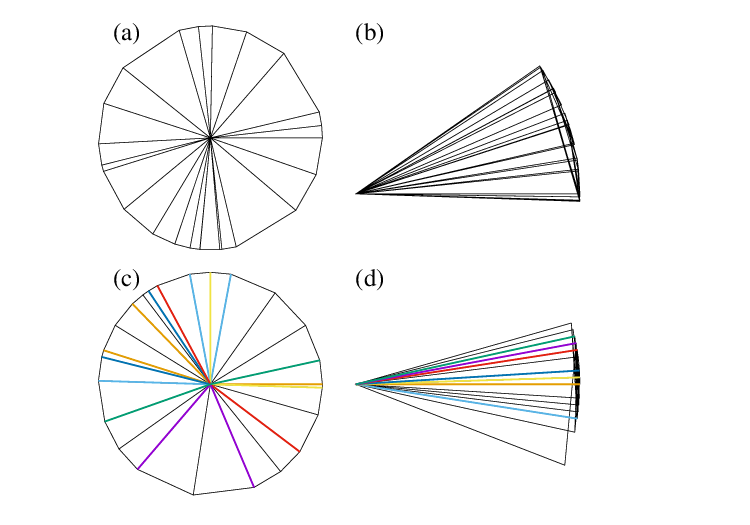}
\caption{Examples of of single vertex origami diagrams,
(a) An origami diagram with $n=24$ and randomly generated width of angles around a center vertex,
(b) A Corresponding pre-folded diagram for the diagram (a),
(c) An origami diagram with $n=24$ and $w=24$.
(d) A Corresponding pre-folded diagram for the diagram (b),
\label{fig:GenModel_Geom}
}
\end{SCfigure}

The instances are generated with two method, the examples of diagrams are shown in Fig.\ref{fig:GenModel_Geom}.
In the first method, the width of each angle around the center is ramdomly given from continuous real numbers in the range $0$ to $1$ by uniform distribution.
The angles are given so that the alternating sum is $0$, and eventually normalized so that the sum is $2\pi$.
This can be expected to correspond to the infinitesimal limit of the width compared to the case where discrete unit width $w$ is introduced to the angles, which will be described later.

\subsection{Definition of quantities}

\subsubsection{Entropy as Logarithm of Total Number of Flat-Foldings}
In statistical mechanics, a quantity called a partition function $Z(\beta)$ works as a generating function to obtain the expected value of various physical quantities, energy, each variable, or the sum of them at a fixed temperature.
Its definition is the summation of the quantity $\exp(-\beta H)$ over combinations of variables included in the energy function $H$, namely,
$Z(\beta)=\sum_{s_{i,j}=\pm 1}\sum_{s_{i',j'}=\pm 1}\cdots \exp(-\beta H(\{s\}))$,
where the summation symbol over many variables, $\sum_{s_{i,j}=\pm 1}\sum_{s_{i',j'=\pm 1}}\cdots$, means that it is taken over all combinations of the values of the variables of the system.
The $H$ discussed in this paper obviously has a minimum value $0$, at least for the origami diagrams that satisfy the Kawasaki theorem. Therefore, the value of $Z(\beta)$ in the limit of $\beta \to \infty$, which is called zero-temperature limit in physics, is the sum of the value $1$ by the total number of flat-foldings. 
The logarithm of the partition function, $F(\beta)=\beta\log Z(\beta)$, is usually called the free energy.
In particular for the energy function with the minimum value $0$, the zero temperature limit $\lim_{\beta \to \infty} F(\beta)$ is also called zero-temperature entropy or ground-state entropy.
The value of the ground-state entropy is numerically obtained by integrating the expected value of energy at a finite value of $\beta$ using the following formula.
This paper describes the results of an approximate evaluation of this absolute zero entropy using a numerical calculation or probabilistic sampling method based on the Markov chain Monte Carlo method, especially the replica exchange Monte Carlo method\cite{NH_JPSJ}.

In this paper, in order to discuss the average behavior for various folding diagrams, the zero-temperature entropy $\lim_{\beta \to \infty} F(\beta)$ on each instance is further averaged over several instances.
Therefore we represent each instance of origami diagram with the symbol $\Delta$ and write the zero temperature entropy for a diagram $\Delta$ as $S_{tot}^{(\Delta)}$. 
In addition, these quantities averaged over all generated $\Delta$ is represented as $[S_{tot}]=\Big(\sum_{\Delta} S_{tot}^{(\Delta)}\Big) \Big/ \Big(\sum_{\Delta}1\Big) $.

\subsubsection{Number of facets sandwitched}
\label{subsubsec:sandw}
If a certain facet $k$ is sandwiched between two other facets $i$ and $j$, the product $s_{i,k}s_{j,k}$ takes the value $\tau_{ik}\tau_{kj}=-J_{(ik)(kj)}$.
Hence, the number of facets which are sandwitched between $i$ and $j$ is obtained by calculating the following quantity,
\begin{equation}
\label{eq:nm_sdw}
n^{\mathrm(sdw)}_{i,j}=\sum_{k}\frac{1-J_{(ik)(kj)}s_{i,k}s_{j,k}}{2},
\end{equation}
where the running suffix $k$ is taken for every $k$ for which a variable $s$ is given between both of the facets $i$ and $j$ those which form a crease.
However, no variables are introduced for facet pairs that do not have a direct vertical relationship due to their overlapping positions. 
In other words, a facet $l$ that does not have a vertical relationship with either of the two facets $i$ or $j$ is not included in the sum $\sum_{k}$ in Eq. (\ref{eq:nm_sdw}).
Therefore, in the scope of this study, facets that are sandwiched indirectly or hidden are not counted in the number.

After computing $n^{\mathrm(sdw)}_{i,j}$ for all creases $(i,j)$ included in the diagram,
the maximum and minimum of $n^{\mathrm(sdw)}_{i,j}$ included in each flat folding are written as $n_{max}$, $n_{min}$.
In addition, for each value of $n_{max}$ and $n_{min}$, we approximately enumerated (the logarithm of) the number of flat foldings that have such values as the maximum and minimum of $n^{\mathrm(sdw)}_{i,j}$s, $S(n_{max})$ and $S(n_{min})$ resepctively, in the same way as $S_{tot}$.
Furthermore, these values are averaged over instances as well as $S_{tot}$. Those are written as $[S(n_{max})]$ and $[S(n_{min})]$.
In the below section \ref{subsubsec:distr_sandw}, they are written $[S(\nu_{max})]$ and $[S(\nu_{min})]$ with $\nu_{max}=n_{max}/n$ and $\nu_{min}=n_{min}/n$.

The number of facets sandwiched between a crease is also called the crease width. 
Awareness of this issue was introduced in the paper \cite{Stamp0} and a research context called folding complexity has been developed.
It is known that the problem of finding a flat-folding that minimizes the maximum value of crease widths $n_{max}$ in stamp-folding problem, in the single vertex origami-diagram with uniform facets and open boundary, is NP-hard.

In introducing $[ S(n_{max}) ]$ or $[ S(n_{min}) ]$, it must be mentioned that the maximum and minimum values of $n_{max}$ or $n_{min}$ are different for each instance.
There is a gap between the average value taken over only instances where non-zero contribution of $S(n_{max})$ is confirmed and the frequency itself of obtaining instances with foldings whose maximum crease width is $n_{max}$.
If an instance does not have a fold whose maximum crease width is $n_{max}$, its number is $0$ and its logarithm is $-\infty$, which is not suitable for averaging.
To deal with this problem with, the following two quantity is computed,
\begin{align}
[S(n_{max})]=\lim_{\alpha \to 0} \Big[ \log\big\{\alpha+\exp\big(S(n_{max})\big)\big\} \Big],\\
\overline{[S(n_{max})]}=\sum_{\Delta \in D^{(n_{max})}}S^{(\Delta)}(n_{max}) \ \Big/ \ \sum_{\Delta \in D^{(n_{max})}}1,
\end{align}
where 
$D^{(n_{max})}$ is the set of instances that have foldings whose maximum crease width is $n_{max}$.

\subsubsection{Overlap between Two Configurations}
As an indicator of how the solution space corresponding to flat folding is embedded in the variable space expressed by the spin variables, we conducted two independent numerical experiments at sufficiently low temperatures and computed the overlap of the spin variables. 
The configurations sampled from two independent numerical simulations at the same temperature are denoted by spin variables $s_{i,j}$ and $s'_{i,j}$, respectively. 
We calculate the following quantities for these pair of simulations,
\begin{equation}
Q=\sum_{{i,j}}\frac{1-s_{i,j}s'_{i,j}}{2}, \label{eq:def_Q}
\end{equation}
where the summation in Eq. (\ref{eq:def_Q}) is taken over all variables of the system composed of $\{s_{i,j}\}$.
Let the total number of variables be $N$ and $q=Q/N$. If $q=1$, the two configurations are completely the same, and if $0$, they are completely reversed. 
Also, when $q \simeq 0.5$, it means that about half of all variables are reversed, and the two configurations can be said to be almost uncorrelated or ``unrelated''.

A normalized histogram $h_{\beta,n}(Q)$ is obtained from the sequence of $Q$s sampled by a pair of numerical simulations with a certain value of (inverse) temperature $\beta$.
Using $h_{\beta,n}(Q)$, the probability density distribution $P_{\beta}(q)$ is derived as,
\begin{equation}
P_{\beta}(q)=\frac{\sum_{Q=0}^{N} h_{\beta,n}(Q) \chi_{q}(Q/N)}{\sum_{Q=0}^{N} h_{\beta,n}(Q)},
\end{equation}
where $\chi_{q}(Q/N)$ is an indicator function that returens the value $1$ when $Q/N=q$ and returns $0$ when $Q/N \neq q$.
This quantity provides informations about how far apart the sampled configurations in a pair of simulations are from each other in variable space at a given temperature.
At the limit of $\beta \to \infty$, this asymptotes to the structure of the embedding of the solution in the configuration space.
If there is not one or only a small number of configurations that completely satisfy the constraints, then $P_{\beta}(q)$ is sufficiently low and $q=1$, and the physical model in question in this paper is considering that it is symmetric about inversion, there is a sharp peak near $q=0$ and $q=1$. 
Furthermore, if there is another satisfied configurations in a region where the values of some variables are different, a peak similarly occurs in the region of $q$ values corresponding to the difference.
The average over instances are taken as well as the entropies, namely $[P_{\beta}(q)]=\sum_{\Delta}\Big\{P_{\beta}^{(\Delta)}(q)\times 1/\Big(\sum_{\Delta}1\Big)\Big\}$.

\section{Results}
\label{sec:results}
%
\subsection{Average behavior of total numbers of foldings}
The logarithm of the total number of flat-foldings in a origami diagram consisting of facets of uniform length is proportional to $n\log n$. 
Whereas, in diagrams with random facets generated by the procedure explained in Section \ref{subsec:ForInstances}, the entropy value which corresponds to the total number of folds exhibits the behavior shown in Fig.\label{fig:figA}.
In Fig.\label{fig:figA} the values of entropy is proportional to $n$ in average, which means that the dependence of the total number of flat-foldings on $n$ is exponential.
This dependence is qualitatively different from the case with uniform facets.
In the exponential function, whose form is $exp(\gamma n\log n)$, the value of $\gamma$ can be readed as $0.098$ from the figure, even though the figure is the result of a numerical experiment and is an averaged value over various instances.
\label{subsec:total_fold}
\begin{SCfigure}[][ht]
  \centering
  \includegraphics[width=0.5\linewidth]{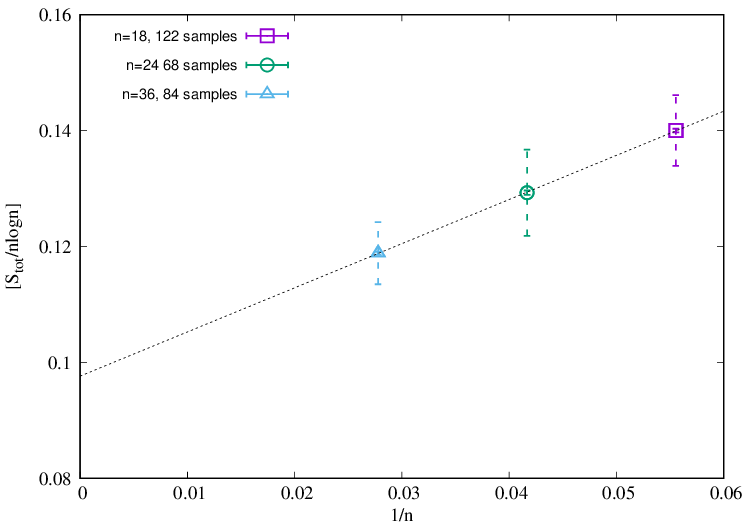}
  \caption{
The values of total entropy  averaged over instances $[S_{tot}]$ is plotted against the inverse of the number of facets, $1/n$, for $n=18$ (square), $24$ (circle) and $36$ (triangle).
Each point is overlaid with an error bar of the average value itself over instances and the variance over instances.
The error bars are given by the bootstrap method and drawn as solid lines, however they are smaller than the size of the mark. 
The variances are written with dashed lines.
}
  \label{fig:figA}
\end{SCfigure}

\subsubsection{Distribution of the minimal and maximal numbers of sandwitched facets}
\label{subsubsec:distr_sandw}
Next, the results regarding the entropy of flat-foldings on the maximum and minimum numbers of facets sandwiched between each crease, respectively noted as $n_{max}$ and $n_{min}$, are described.

\begin{figure}
  \centering
  \subfloat[]{
    \includegraphics[width=50mm]{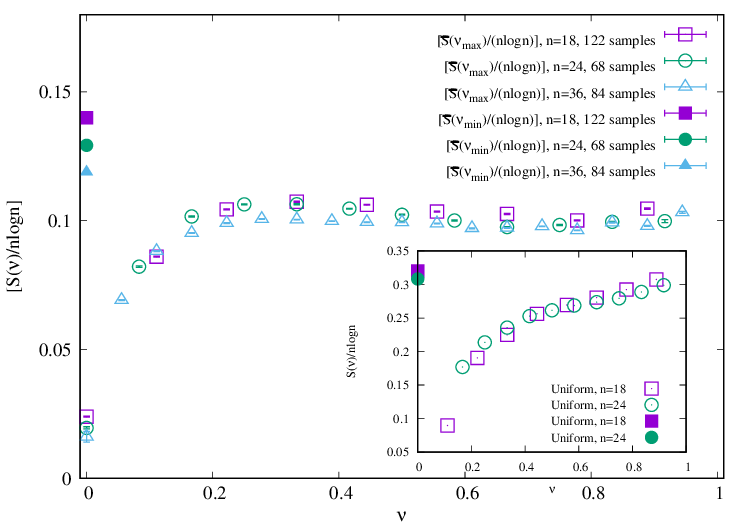}
    \label{fig:subfig:distr_entr_scale}
  }
  \subfloat[]{
    \includegraphics[width=50mm]{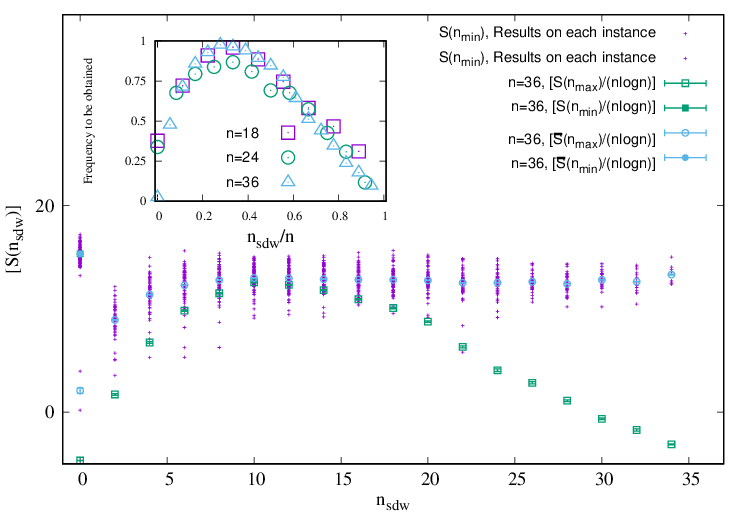}
    \label{fig:subfig:distr_various_def}
  }  
  \label{fig:8OSMElogosubfigs}
  \caption{(a)The values of $\overline{[S(\nu_{max})]}$ and $\overline{[S(\nu_{min})]}$ are plotted against $\nu_{max}$ or $\nu_{min}$ for $n=18$ (square), $24$ (circle) and $36$ (triangle). The mark is open for $\overline{[S(\nu_{max})]}$ and closed for $\overline{[S(\nu_{min})]}$. 
In the inset the values $S(\nu_{max})$ and $S(\nu_{min})$ for diagrams of uniform angle width, for comparison. (b)Comparison between $[S(n_{max})]$ (square) and $\overline{[S(n_{max})]}$ (circle) with $n=24$. In the inset the frequency of obtaining instances with foldings whose maximum crease width is $n_{max}$ for $n=18$ (square), $24$ (circle), and $36$ (triangle).}
\end{figure}

In Figure \ref{fig:subfig:distr_entr_scale} the horizontal axis describes $n_{max}$ or $n_{min}$ divided by the total number of facets of the system $n$,
which are respectively represented as $\nu_{max}=n_{max}/n$ and $\nu_{min}=n_{min}/n$,
 and the vertical axis is the logarithm of the total number of the flat foldings such that the value of $n_{max}$ or $n_{min}$ takes the value on the horizontal axis, $[ S(\nu_{max}) ]$ and $[ S(\nu_{max}) ]$ with respect devided by $n \log n$.
A curve is shown for $n=18,24$ and $36$.
For reference, the same quantities for a single vertex diagram with uniform facet angle are shown in the inset.

The same as $[S_{tot}]$, the curves of $[S(\nu_{max})]$ or $[S(\nu_{min})]$ for various sizes appears to collapse into a master curve when divided by $n \log n$.
The master curve is found in both cases with uniform and random facets. 
However, those two exhibit qualitatively different behaviors.
In the case of uniform facets, the curve of $S(n_{max})$ increases monotonically as the value of $n_{max}$ increases. 
This suggests that various combinations of stacking orders are realized even among the facets sandwiched between the creases with the maximum width.
On the other hand, in the case of random facets, the value of $S(\nu_{max})$ shows a behavior that decreases slightly in the region where the value of $n_{max}$ is large.
At $n_{max}=n-2$, it roughly takes the maximum value of $S(\nu_{max}) \simeq 0.3 \times n\log n$. 
Whereas, in the case of random angle width $[S(\nu_{max})]$ takes its maximum value, approximately $\overline{[S(\nu_{max})]} \simeq 0.1 \times n\log n$, at $\nu_{max} \simeq 0.3$ and decreases once the value of $\nu_{max}$ exceeds that argument as shown in Fig.\ref{fig:subfig:distr_entr_scale}.
Also regarding $\overline{[S(\nu_{max})]}$, this scaling behavior, including the value of the coefficient, of the maximum value is the same as shown in Fig.\ref{fig:subfig:distr_various_def}, although the decrease of the value is rather significant in this case bacause the fraction becomes smaller as $n_{max}$ goes to its large value.

Meanwhile, $S(\nu_{min})$ or $[S(\nu_{min})]$ have its contribution only on $\nu_{min}=0$, both for origami diagrams with random and uniform angle width.
This is thought to be due to the fact that in the single-vertex diagram the array of facets is closed, that is, all facets have creases at both side of them and no have an open end.
In the case with exact stamp folding problem, there obviously exists a folding in which the minimum value of the crease width is a finite positive value.

Note here that there is nothing unnatural about the curve of $[S(\nu_{max})]$ having a finite number of folds assigned to the position $\nu_{max}=0$.
For example, if a random folding diagram contains two facets with fairly large angles and they are not connected by a crease, it is possible to create a folding where one of the two is located at the top in the vertical relationship of the overall stacking and the other is at the bottom.
In such a folding, between the top and bottom facets, two stacks may be formed at a considerable distance from each other.
Due to the circumstances described in Section \ref{subsubsec:sandw}, in such stacking there may be the case in any creases and two facets that compose them no other facets sandwiched between.

\subsubsection{Averaged Shape of Overlap Distribution}
\label{subsec:distr_ham}
\begin{figure}
  \centering
  \subfloat[]{
    \includegraphics[width=50mm]{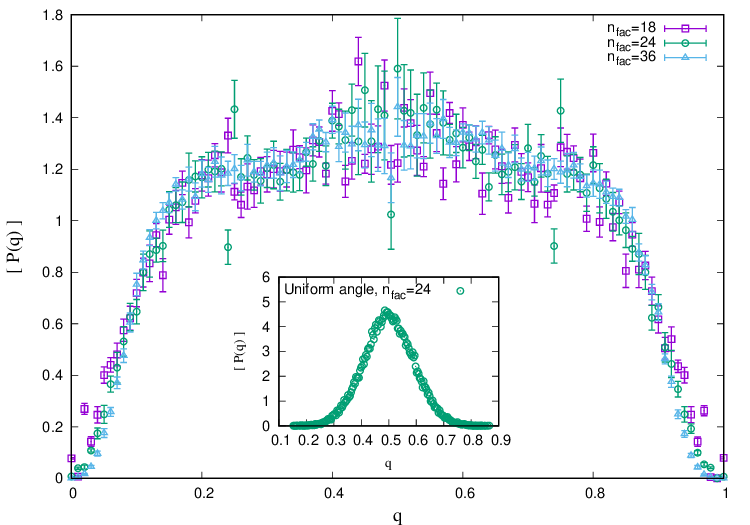}
    \label{fig:ovlp_d}
  }
  \subfloat[]{
    \includegraphics[width=50mm]{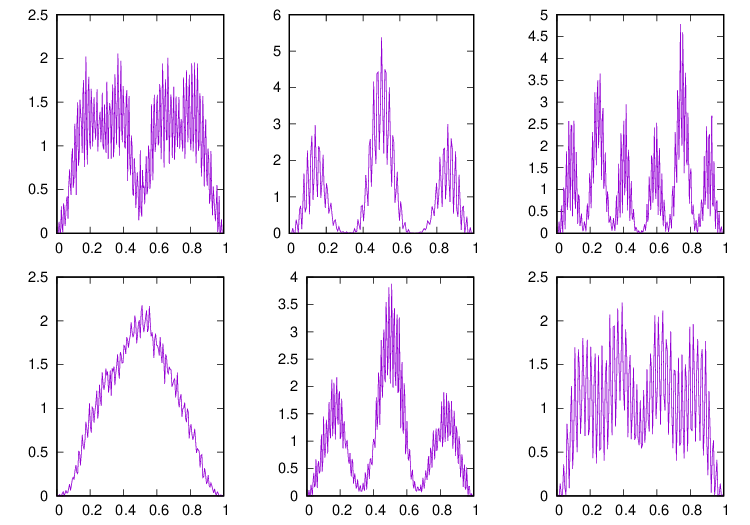}
    \label{fig:ovlp_d_each}
  }  
  \caption{(a)The instance-averaged probability density function $[P_{\beta}(q)]$ plotted against the value of $q$ for $n=18$ (square), $24$ (circle) and $36$ (triangle). In the inset, $P_{\beta}(q)$ for the origami diagram with uniform angle width is shown for comparison. (b)Examples of $P_{\beta}(q)$ of each instances with $n=24$.}
\end{figure}

The results of this study are shown in Figure \ref{fig:ovlp_d}.
The contribution of the distribution is widely dispersed, with the ratio of the degree of overlap to the total number of variables $q$ ranging from about $0.2$ to $0.8$. 
Comparing to the shape of the overlap distribution of the origami diagram with a uniform angular width is a Gaussian,
 that on diagrams with random angular widths clearly has the different shape. 
Instead, the contribution of $[P(q)]$ is large even in regions where the absolute value of $q$ is far from $0$. 
Therefore, it can be seen that the set of flat foldings is composed of stacks of facets whose vertical relationships are very different from each other.

As a consequence of the theory of replica symmetry breaking in statistical mechanics, the three peak structure in the asymptotic shape of $[P(q)]$ (at $q=0.5$ and symmetrical positions centered on it) implies that the average complexity class of the seach problem is NP-complete\cite{MezMon}.
When the size dependence shown in Fig. \ref{fig:ovlp_d} is focused based on this viewpoint, the shape found in $0.2<q<0.8$ in the range up to $n = 36$ somewhat seems to lead the three peak structure.
Meanwhile,the width of the error bar is also large. 
Thus, it yet require the carefull observation to conclude that the average-case computational complexity of searching flat foldings of random single-vertex origami diagrams is NP-complete.

\subsection{Reduction of System Size for Computational Efficiency}
In the physical model given by Eq. (\ref{eq:tot_en_func}), each term with the form (\ref{eq:term_J}) in particular gives a relationship that should be satisfied in the ground state between variables respectively involved in each term.
Applying this relationship to a set of variables in (P) and rearranging them brings an efficiency for numerical approximations to the total number of flat folds and overlapping distributions we have seen in the above by mapping the ground state to a system described by fewer variables.
In addition, this procedure is expected to aggregate variables whose values are determined uniquely each other into a single cluster and to reveal the combinatorial structure that is the essential cause of the problem.

\begin{figure}
  \centering
  \subfloat[]{
    \includegraphics[width=50mm]{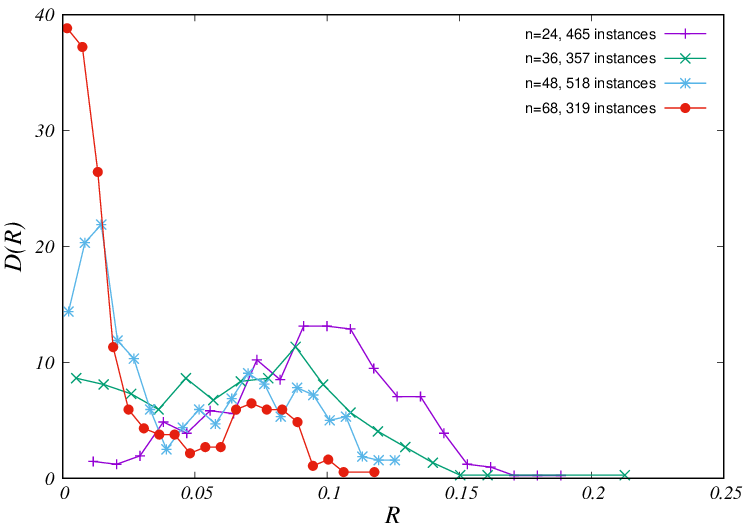}
    \label{fig:DistributionInRandom}
  }
  \subfloat[]{
    \includegraphics[width=50mm]{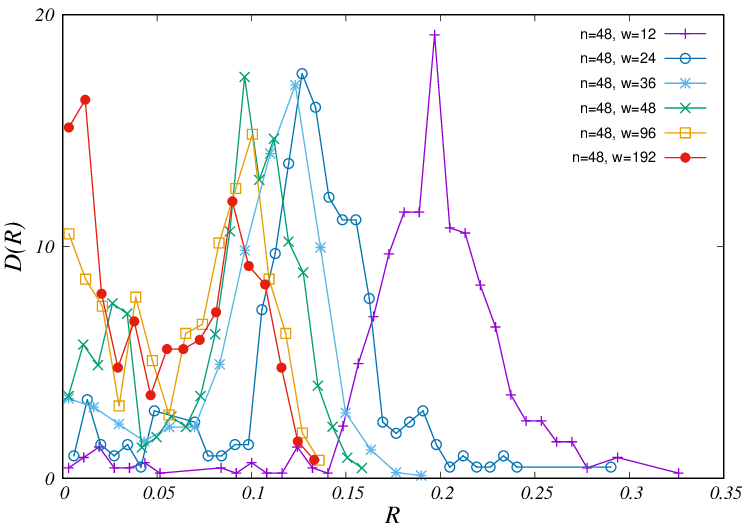}
    \label{fig:Distribution_wdep}
  }  
  \caption{(a)Density plot of the reduction ratio, $D(R)$ for various instances with randomly and continuously generated width of each angle.
Dependence on values of $n=36, 48,$ and $68$ with randomly generated values of angles. 
(b)Dependence of $D(R)$ on various values of parameter $w=12, 24, 36, 48, 96, 128$ with fixed value $n=48$.}
\end{figure}

Here, the result of the reduction of variables by the contraction via the equations (\ref{eq:term_J}) for the origami diagram instances ranging from $n=24$ to $68$.
For the origami diagram obtained using the method in Sec.\ref{subsec:ForInstances}, the variables are reduced as demonstrated in Sec. \ref{subsec:Contract}.
This reduced number of variables is represented as $\tilde{N}$ and the ratio $R=\tilde{N}/N$ is discusssed for each instance.
Note that there are many cases that the reduction result as the collection of several independent component each consisting of variables $\{s,C\}$. 
The two variables $s_{i,j}$ or $C_{l}$ which belong to each different component are not involved in a same product term.
For such cases let $\tilde{N}$ be the number of variables contained in the largest component in the collection.
The density plot of the distribution of the ratio $D(R)$ for $s = 24,48,68$, from $465$, $518$, $319$ instances respectively, is shown in Fig.\ref{fig:DistributionInRandom}. 
In Fig.\ref{fig:DistributionInRandom}, for small $n$, for example $n=24$, the distribution has a unimodal peak roughly around $0.05<R<0.15$.
However, in the case of a larger system size, another peak appears on the side with a smaller value of $R$.
As the system size further increases, the distribution becomes bimodal, with the height of two peaks swapping, and the peak with the smaller $R$ value becoming dominant.

A similar change in the shape of $D(R)$ also occurs in the case with sets of instances generated by the second method, the method with the minimum unit is introduced to the angular width.
As shown in Fig.\ref{fig:Distribution_wdep}, it is observed when the value of $s$ is fixed and the values of $w$ are increased.
From this observation, it is thought that when the number of terms with a product of four-variables like Eq.(\ref{eq:terms_K}) is large, $D(R)$ has a unimodal shape with its peak in the region of relatively large values of $R$,
and as the number decreases the shape changes to one with a dominant peak on the small-$R$ side via the bimodal shape.

However, if the angular width takes continuous random values for a small $n$, the number of terms like Eq.(\ref{eq:terms_K}) in the corresponding cost function still remains almost $0$, which does not mean that there are a large number of terms.
For this point further consideration on the relationship to the number of the four-variables product is currently required.\\
Within the range of system sizes tested, the system size $\tilde{N}$ can be reduced to roughly $1/20$ of the original system size.
Compared to this, in the diagram with a uniform angles width, we have to treat the system with $n(n-1)/2$ variables because there is no room for reducing the variables using the contraction procedure.
This suggests that the average computational amount for a set of random instances can behave quite differently compared to that for a uniform angular value, which is considered to be the corresponding to the worst-case computational complexity.
It is also interesting from a computational complexity perspective.

\subsubsection{Trying to obtain system-size dependence of each separated peak}
In Fig. \ref{fig:DistributionInRandom}, it is found that the peak on the small $R$ side in $D(R)$ gradually becomes more prominent as the larger $n$. 
Although being aware that this may not be an accurate evaluation, an attempt to evaluate the asymptotics of this peak is made as shown in Fig.7. 
The dispersion $\sigma_R$ of the small-$R$ side peak, meaning the width of the peak, is computed using only instances whose $R$ are less than or equal to a cutoff.
\begin{SCfigure}
\includegraphics[width=50mm]{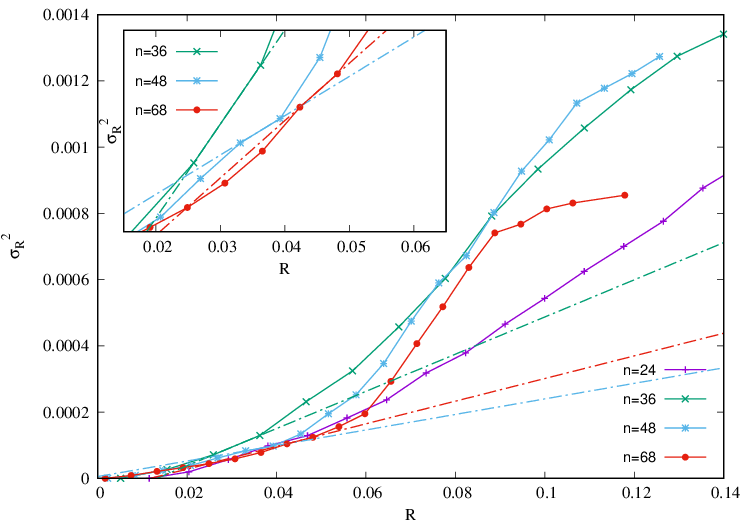}
\caption{
The dispersion $\sigma_R$ for the lower side peak in $D(R)$ with various thresholding values.
\label{fig:Dep_Dispersion_InRandom}
}
\end{SCfigure}
While the cutoff is located between the two peaks, the slope of the variance change is relatively small, and when the cutoff is located on the small $R$ peak, the slope of the change is large. 
Using this behavior, we estimated the value at which the slope of the curve representing dependence begins to increase (again) as seen in the caption of Fig.7. 
The threshold value of $R$ is estimated as $Rt = 0.030, 0.037, 0.045$ for $n = 36, 48, 68$, respectively. 
These values exhibit that it is also not possible to exclude the possibility that $[\tilde{N}]$ has a depencence of sub-linear order on $[N]$, where $[\tilde{N}]$ and $[N]$ is averaged over instances because the system size $N$ is different among each instances.

\section{Summary}
\label{sec:summary}
In this study based on the physical model formulation, approximate enumeration of the total number of foldings and its decomposition into those with fixed maximum and minimum values of the number of facets between each crease, i.e., the crease width.
The value of the logarithm of the total numbers of the configuration, computed by the replica exchange Monte Carlo method, is $0.098 \times n\log n$ and rather smaller than that regarding the origami diagrams with uniform angle width.
Furthermore, the behaviors of the decomposed entropy with the maximum number of facets, $n_{max}$ sandwithed by each crease of origami diagrams of respectively random or uniform angle width behave qualitatively different from  each other.
This comparison implys that the constraints on mutual penetration of facets sandwiched within the crease with the maximum width brings a limit of the diversity of combinations of their stacking orders.

In addition, we approach the average computational complexity of the problem of enumerating the number of flat folding of the single vertex origami diagram from two perspectives: observation of the behavior of overlap distribution functions from the viewpoint of replica symmetry breaking in spin glass theory, and reduction of system size based on chain relationships that fix the values of variables.
Given the range of system sizes observed in this study, it has still not been possible to draw conclusions regarding the results from both and their consistency with each other. However, it is expected that research in this direction will still have the potential to be a future topic.
%

\section{Appendix}
\label{sec:appendix}
%

\subsection{Contraction of spin variables}
\label{subsec:Contract}
\begin{SCfigure}
\includegraphics[width=50mm]{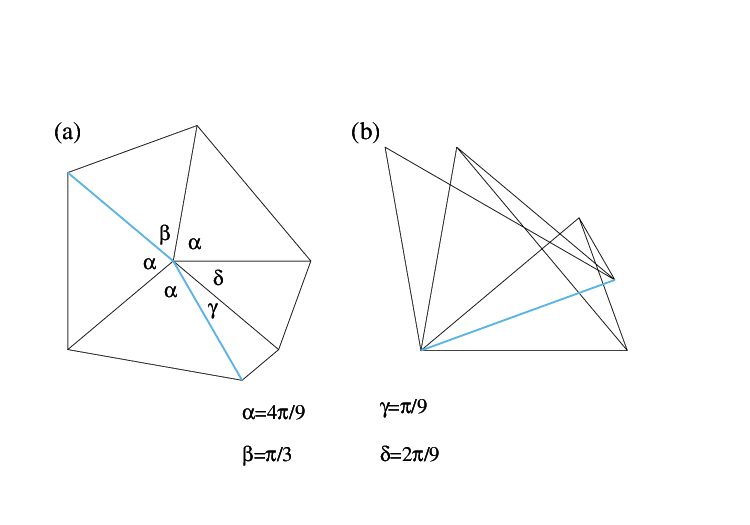}
\caption{
An example of diagram of single vertex origami with $n=6$. $\alpha$, $\beta$, $\gamma$, $\delta$ at the bottom of the figure are the angle values around the center point of facets.
 two colored creases in (a) overlap when the diagram is flat-folded as in (b).\label{fig:SmallSample}}
\end{SCfigure}

Here, the method of variable reduction is intruduced based on the example of the origami diagram shown in Fig.\ref{fig:SmallSample}.
First, from the diagram of this figure, the following cost function is obtained by the modeling described in Section \ref{sec:physmodel}.
In a term whose form is the same as Eq.(\ref{eq:term_J}), the combination of the involved spin variables that brings the value $0$ of the term is uniquely determined except for total inversion of the both two.
Therefore, we can translate the constraint from the term like Eq.(\ref{eq:term_J}) into an allowed relationship between the spin variables.
By using such translation, several spin variables that appear in the energy function can be collectively re-expressed as a single variable. 
And the energy function can be re-given as a combination of fewer binary variables.
\label{eq:cf_ex}
\begin{align}
\sum_{(ijkl)}E^{(q)}_{ijkl}&=\frac{1+s_{24}s_{25}s_{34}s_{35}}{2},\label{eq:terms_K_Ex}\\
\sum_{(ij,k)}E^{(i)}_{ij,k}&=\frac{1-s_{13}s_{23}}{2}+\frac{1-s_{14}s_{24}}{2}+\frac{1-s_{12}s_{13}}{2}+\frac{1-s_{26}s_{36}}{2} \nonumber\\
\label{eq:terms_J_Ex}
&\qquad+\frac{1-s_{14}s_{15}}{2}+\frac{1-s_{46}s_{56}}{2}+\frac{1-s_{15}s_{16}}{2}+\frac{1-s_{25}s_{26}}{2} \nonumber\\
&\qquad+\frac{1-s_{35}s_{36}}{2}+\frac{1-s_{45}s_{46}}{2},\\
\label{eq:terms_L_Ex}
\sum_{(ijk)}E^{(c)}_{ijk}&=\frac{1+s_{12}s_{23}-s_{23}s_{13}-s_{13}s_{12}}{4}+\cdots+\frac{1+s_{35}s_{56}-s_{56}s_{36}-s_{36}s_{35}}{4},
\end{align}
where the detail of Eq.(\ref{eq:terms_L_Ex}) is written as Eq.(\ref{eq:dtl_trip_top})-(\ref{eq:dtl_trip_bottom}), described later.
By applying the rewriting process based on the product of two spin variables, which is mentioned at the beginning of this section, from the 2nd to 11th term of the above cost function (\ref{eq:cf_ex}) each leads the relationship between two variables.
The relationships are eventually summarized as follows,
\begin{align}
C_1=s_{12}=s_{13}=s_{23},\\
C_2=s_{14}=s_{24}=s_{15}=s_{16},\\ 
C_3=s_{26}=s_{36}=s_{25}=s_{35},\\
C_4=s_{45}=s_{46}=s_{56}.
\end{align}
With the above introduced cluster variables, the terms of Eq.(\ref{eq:terms_L_Ex}) is rewritten as
\begin{align}
\label{eq:dtl_trip_top}
\frac{1+s_{12}s_{25}-s_{25}s_{15}-s_{15}s_{12}}{4}&=\frac{1+C_1C_3-C_3C_2-C_2C_1}{4},\\
\frac{1+s_{12}s_{26}-s_{26}s_{16}-s_{16}s_{12}}{4}&=\frac{1+C_1C_3-C_3C_2-C_2C_1}{4},\\
\label{eq:srce_rw_b1}
\frac{1+s_{34}s_{45}-s_{45}s_{35}-s_{35}s_{34}}{4}&=\frac{1+s_{34}C_4-C_4C_3-C_3s_{34}}{4},\\
\label{eq:srce_rw_b2}
\frac{1+s_{34}s_{46}-s_{46}s_{36}-s_{36}s_{34}}{4}&=\frac{1+s_{34}C_4-C_4C_3-C_3s_{34}}{4},\\
\frac{1+s_{12}s_{23}-s_{23}s_{13}-s_{13}s_{12}}{4}&=\frac{1+C_1C_1-C_1C_1-C_1C_1}{4}=0\\
\frac{1+s_{12}s_{24}-s_{24}s_{14}-s_{14}s_{12}}{4}&=\frac{1+C_1C_2-C_2C_2-C_2C_1}{4}=0 \label{eq:srce_rw_a}\\
\frac{1+s_{13}s_{34}-s_{34}s_{14}-s_{14}s_{13}}{4}&=\frac{1+C_1s_{34}-s_{34}C_2-C_2C_1}{4},\\
\frac{1+s_{13}s_{35}-s_{35}s_{15}-s_{15}s_{13}}{4}&=\frac{1+C_1C_3-C_3C_2-C_2C_1}{4}\\
\frac{1+s_{13}s_{36}-s_{36}s_{16}-s_{16}s_{13}}{4}&=\frac{1+C_1C_3-C_3C_2-C_2C_1}{4},\\
\frac{1+s_{14}s_{45}-s_{45}s_{15}-s_{15}s_{14}}{4}&=\frac{1+C_2C_4-C_4C_2-C_2C_2}{4}=0,
\end{align}
\begin{align}
\frac{1+s_{14}s_{46}-s_{46}s_{16}-s_{16}s_{14}}{4}&=\frac{1+C_2C_4-C_4C_2-C_2C_2}{4}=0,\\
\frac{1+s_{23}s_{35}-s_{35}s_{25}-s_{25}s_{23}}{4}&=\frac{1+C_1C_3-C_3C_2-C_2C_1}{4}, \\
\frac{1+s_{23}s_{36}-s_{36}s_{26}-s_{26}s_{23}}{4}&=\frac{1+C_1C_3-C_3C_2-C_2C_1}{4},\label{eq:srce_rw_b3}\\
\frac{1+s_{24}s_{45}-s_{45}s_{25}-s_{25}s_{24}}{4}&=\frac{1+C_2C_4-C_4C_3-C_3C_2}{4},\label{eq:srce_rw_b4}\\
\frac{1+s_{24}s_{46}-s_{46}s_{26}-s_{26}s_{24}}{4}&=\frac{1+C_2C_4-C_4C_3-C_3C_2}{4},\\
\frac{1+s_{15}s_{56}-s_{56}s_{16}-s_{16}s_{15}}{4}&=\frac{1+C_2C_4-C_4C_2-C_2C_2}{4}=0, \\
\frac{1+s_{45}s_{56}-s_{56}s_{46}-s_{46}s_{45}}{4}&=\frac{1+C_4C_4-C_4C_4-C_4C_4}{4}=0,\\
\frac{1+s_{25}s_{56}-s_{56}s_{26}-s_{26}s_{25}}{4}&=\frac{1+C_3C_4-C_4C_3-C_3C_3}{4}=0,\label{eq:dtl_trip_bottom}\\
\frac{1+s_{35}s_{56}-s_{56}s_{36}-s_{36}s_{35}}{4}&=\frac{1+C_3C_4-C_4C_3-C_3C_3}{4}=0.
\end{align}

In particular, we focus on the rewriting of the first term, Eq.(\ref{eq:terms_K_Ex}), which is
\begin{equation}
\label{eq:Regen_J_Ex}
\frac{1+s_{24}s_{25}s_{24}s_{35}}{2}=\frac{1+C_2C_3s_{34}C_3}{2}=\frac{1+C_2s_{34}}{2},
\end{equation}
whose form composes the product of two spin variables similar to Eq.(\ref{eq:term_J}) again.
The reproduction of the term whose form is similar to Eq.(\ref{eq:term_J}) induces the recursive application of the rewriting process.
Here, the following relationship,
\begin{equation}
\label{eq:rewrite_mid}
s_{34}=-C_2,
\end{equation}
is lead from Eq.(\ref{eq:Regen_J_Ex}).
In addition, further rewritings and relationships are induced from Eq.(\ref{eq:rewrite_mid}) as follows,
\begin{align}
\label{eq:rwt_st2_a}
\frac{1-C_1C_2+C_2C_2-C_2C_1}{4}&=\frac{1-C_1C_2}{2},\\ 
\label{eq:rwt_st2_b}
\frac{1-C_2C_4-C_4C_3+C_3C_2}{4}+
\frac{1+C_2C_4-C_4C_3-C_3C_2}{4}&=\frac{1-C_4C_3}{2},
\end{align}
which results
\begin{align}
\label{eq:rls_st2_a}
C_1=C_2,\\
\label{eq:rls_st2_b}
C_3=C_4,
\end{align}
where Eqs(\ref{eq:rwt_st2_a}) and (\ref{eq:rwt_st2_b}) are brought from Eqs.(\ref{eq:srce_rw_a}) and (\ref{eq:srce_rw_b1})$+$(\ref{eq:srce_rw_b2})$+$(\ref{eq:srce_rw_b3})$+$(\ref{eq:srce_rw_b4}), respectively.

The method of variable reduction explained here utilizes the relationship between variables, meaning the vertical relationship of facets, that must be satisfied in the ground state (state of zero energy) of the physical model. 
The concept that gives a cascading chain of decisions on the vertical relationship of facets is also studied in mathematics and information science, and is called a forcing set.
Algorithms for finding forcing sets for one-dimensional origami\cite{DDDF} and Miura-ori diagrams\cite{BDEFGH} have been proposed, however it appears that no algorithms exist for general origami at present.
Based on a series of studies on this subject, the original definition of a forcing set is considered to be the subset of minimal assignments of the crease patterns that determine the assignments for all creases included in the origami diagram. 
It is known that the problem of giving an entire forcing set is can be an NP-complete problem.
However, a chain of decisions for stacking relationships can be given for local relationships among facets. 
Therefore, it is an interesting application to make the determination of the possibility or enumeration of flat-folding more efficient by partially utilizing the subsets of chained decisions contained in forcing sets.
The method demonstrated in this section is expected to provide insight into the derivation of forced sets for general origami.

\subsection{Numerical method for approximately estimating the number of folding}
\label{subsec:ForEstimation}
The numerical simulations performed for this study are based on the replica-exchange Monte Carlo method and the multiple histogram reweighting technique.

Each variable $s_{i,j}$ is updated using the Metropolis rule. That is, the accepting probability for local updating,  $p_{ud}$, is given by
\begin{equation}\label{eq:p_ud}
p_{ud}=\mathrm{min}\{1,\exp\big(-\beta (E'-E)\big)\},
\end{equation}
where $\beta$ is the physical inverse temperature, $E$ and $E'$ is the value of Hamiltonian (\ref{eq:tot_en_func}) for configuration respectively before and after the update.
The exchange of replicas with indices $l$ and $l+1$ is performed using the Metropolis rule with excahge probability $p_{exch}$, which is given as
\begin{equation}\label{eq:p_exch}
p_{exch}=\mathrm{min}\{1,\exp\big((\beta^{(l)}-\beta^{(l+1)})(H^{(l+1)}-H^{(l)})\big)\}.
\end{equation}

The histograms sampled with each replica, $h_l(E,Q)$, are integrated to estimate the number of states $W(E,Q)$ with the multiple histogram reweighting method.
$W(E,Q)$ is obtained via the equation
\begin{equation}\label{eq:recursion}
W(E,Q)=\frac{\sum_{l}h_l(E,Q)}{\sum_{l}\big(\frac{\omega_l(E,Q)}{z_l}\sum_{E=0}^{n}\sum_{Q=0}^{n}h_{l}(E,Q)\big)},
\end{equation}
where $\omega_l(E,Q)=\exp\big(-\beta_l E\big), \label{eq:omega}$ and $z_l=\sum_{E=0}^{n}\sum_{Q=0}^{n}W(E,Q)\omega_l(E,Q)$.\\
Eq. (\ref{eq:recursion}) is originally derived in the paper 
\cite{FS_PRL}.

\bibliographystyle{osmebibstyle}
\bibliography{biblio_85}

\theaffiliations

\end{document}